\documentclass[aps,twocolumn,amssymb,prl,showpacs,10pt]{revtex4-1}

\usepackage[utf8]{inputenc}
\usepackage{tabularx}
\usepackage{multirow}
\usepackage{url}
\usepackage{amsmath}
\usepackage{graphicx}
\usepackage{mathrsfs}
\usepackage{xcolor}
\usepackage[normalem]{ulem}
\usepackage{float}
\usepackage{lmodern}
\usepackage{soul}
\usepackage[breaklinks,colorlinks,urlcolor=blue,citecolor=blue]{hyperref}

\newcommand{\apjl}{The Astrophysical Journal Letters}%
\newcommand{\mnras}{MNRAS}%

\renewcommand{\vec}[1]{\mathbf{#1}}
\newcommand{\nvec}[1]{\hat{\vec{#1}}}

\def\bB{\boldsymbol{B}}

\newcommand{\refdel}[1]{}

\newcommand{\psection}[1]{\textit{#1.---}}

\begin{document}

\title{
Heating of Magnetically Dominated Plasma by Alfv\'en-Wave Turbulence
}

\author{Joonas N\"{a}ttil\"{a}$^{1,2}$}\email{jnattila@flatironinstitute.org} 
\author{Andrei M.\,Beloborodov$^{1,3}$}

\affiliation{$^1$Physics Department and Columbia Astrophysics Laboratory, Columbia University, 538 West 120th Street, New York, New York 10027, USA}
\affiliation{$^2$Center for Computational Astrophysics, Flatiron Institute, 162 Fifth Avenue, New York, New York 10010, USA}
\affiliation{$^3$Max Planck Institute for Astrophysics, Karl-Schwarzschild-Strasse 1, D-85741, Garching, Germany}

\received{2 August 2021}
\revised{23 November 2021}
\accepted{8 December 2021}

\begin{abstract}
Magnetic energy around astrophysical compact objects can strongly dominate over plasma rest mass. 
Emission observed from these systems may be fed by dissipation of Alfv\'en wave turbulence, which cascades to small damping scales, energizing the plasma. 
We use 3D kinetic simulations to investigate this process. 
When the cascade is  excited naturally, by colliding large-scale Alfv\'en waves, we observe quasithermal heating with no nonthermal particle acceleration. 
We also find that the particles are energized along the magnetic field lines and so are poor producers of synchrotron radiation. 
At low plasma densities, our simulations show the transition to ``charge-starved'' cascades, with a distinct damping mechanism.
\end{abstract}

\maketitle

\psection{Introduction}
Plasmas around neutron stars and black holes are often collisionless and strongly magnetized.
Magnetic energy density in these systems can far exceed the plasma rest mass and provide an ample energy reservoir for heating the plasma and making it a bright source of radiation.
Magnetic dissipation is likely generic across various astrophysical systems including
jets from black holes \citep{Bottcher_2019}, magnetars \citep{Kaspi_2017}, pulsars and their winds \citep{Buhler_2014}, and mergers of compact objects \citep{Ascenzi_2021}.

One possible mechanism for magnetic energy release is the dissipation of Alfv\'en wave turbulence in the magnetically dominated plasma, also known as relativistic turbulence \citep{Thompson_1998, Uzdensky_2018, Sobacchi_2019b, Li_2019}.
Recent numerical experiments demonstrated that when an initial magnetic field $\vec{B}_0$ is stirred by a strong external perturbation, a turbulence cascade develops that can accelerate nonthermal particles 
\citep{Zhdankin_2017a, Comisso_2018, Comisso_2019, Nattila_2021, Zhdankin_21, Hankla_21, Vega_21}.
These works used rather violent excitation of the turbulence by driving external electric currents into the plasma or by starting with strongly deformed magnetic fields, which both result in $\delta B \sim B_0$.

In this Letter, we investigate relativistic turbulence excited by collisions of low-frequency Alfv\'en waves with amplitudes $\delta B/B_0<1$.
This canonical mechanism of turbulence development involves nonlinear interactions between the waves \citep{Goldreich_1995, Troischt_2004, Luo_2006, Thompson_2008, Cho_2014a, Howes_2013, Groselj_2019, Tenbarge_2021}.
Low-frequency Alfv\'en waves are natural initial magnetohydrodynamic (MHD) perturbations, which are easily excited by shear motions in the system. 
The perturbations are long-lived \citep{Mallet_2021} and ducted along $\vec{B}$, which promotes multiple wave collisions. 
Below we present the first, large-scale, fully kinetic, 3D simulations of the  relativistic turbulence generated by this mechanism.
Such simulations allow one to study from first principles the formation of the cascade, eventual dissipation of the waves, and the resulting conversion of magnetic energy to particle energy.

\psection{Simulation setup}
The plasma in our simulations is made of electrons and positrons. 
It is initially cold with a dimensionless temperature $\theta_0 \equiv k_\mathrm{B} T_0/m_\mathrm{e} c^2 = 10^{-5}$ (where $k_\mathrm{B} T_0$ is the initial temperature, $m_\mathrm{e}$ is the electron rest mass, $c$ is the speed of light) and has a uniform density $n_0$.
The problem is not sensitive to the initial temperature: 
we have also tested $\theta_0 = 0.3$ and obtained similar results.
A uniform background magnetic field $\vec{B}_0 = B_0 \nvec{z}$ defines the magnetization $\sigma_0 \equiv  B_0^2/4\pi n_0 m_\mathrm{e} c^2\gg 1$. 
We have studied configurations with $\sigma_0 = 30$, $100$, and $1000$.

We initiate the simulation with two periodic, plane-Alfv\'en waves superimposed on top of each other and $\vec{B}_0$ (see also \citep{Howes_2013, Nielson_2013}):
\begin{equation}\begin{split}\label{eq:deltab}
    \delta \vec{B}_1 &= -\delta B \cos(k_{\perp,0} y + k_{\parallel,0} z ) ~\nvec{x}, \\
    \delta \vec{B}_2 &= +\delta B \cos(k_{\perp,0} x - k_{\parallel,0} z ) ~\nvec{y}.
\end{split}\end{equation}
The electric fields of the waves are $\delta \vec{E}_{1,2} = v_\mathrm{A} \nvec{z} \times \delta \vec{B}_{1,2}/c$ where $v_\mathrm{A}/c \equiv \sqrt{\sigma_0/(\sigma_0+1)}\approx 1$ is the Alfv\'en speed.
The waves have two dimensionless parameters: amplitude $\delta B/B_0$ and obliqueness $k_{\perp,0}/k_{\parallel,0}$.
In MHD, nonlinear interaction is activated for waves with nonaligned polarizations \citep{Howes_2013, Tenbarge_2021, Note1strong};
we use $\delta\vec{B}_1\perp \delta\vec{B}_2$. 
The two waves propagate in the opposite directions $\pm \nvec{z}$ and cross one (parallel) wavelength on the timescale $t_\parallel = 2\pi/ck_{\parallel,0}$.
We also experimented with more elaborate initial configurations (e.g., colliding Alfv\'en wave packets \cite{Verniero_2018, Li_2019, Ripperda_2021}), with similar results. 
We chose the simplest configuration in Eq.~\eqref{eq:deltab}, because its evolution well captures the development of turbulence cascade and plasma energization that we wish to study.

The strength of the nonlinear interaction is determined by the parameter,
\begin{equation}
    \chi_0 \equiv  \frac{2\delta B}{B_0} \frac{k_{\perp,0}}{k_{\parallel,0}},
\end{equation}
where $2 \delta B=\delta B_1 + \delta B_2$.
Our fiducial model presented below has $\delta B/B_0 = 1/4$ and $k_{\perp,0}/k_{\parallel,0} = 4$, resulting in $\chi_0=2$ .

\psection{Numerical method}
We use a 3D simulation domain $L_\perp\times L_\perp \times L_\parallel$ with $L_\perp=2\pi/k_{\perp,0}$ and $L_\parallel=2\pi/k_{\parallel,0}$, which satisfies periodic boundary conditions.
Our fiducial simulation has the grid size $1280^2\times5120$.
The characteristic plasma scale $c/\omega_\mathrm{p}$ (where $\omega_\mathrm{p}^2 = 4\pi e^2 n_0/m_\mathrm{e}$) is resolved with 6 cells. 
The initial wave number $k_{\perp,0}\approx  0.03 \, \omega_\mathrm{p}/c$ is well separated from the plasma scale, allowing a significant range for the turbulence cascade.
The simulation is run until most of the turbulence energy becomes dissipated; 
e.g., the run time for our fiducial simulation with $\chi_0=2$ is $10\,t_\parallel$.

The self-consistent evolution of the plasma and the electromagnetic field is followed using the open-source code \textsc{runko}  \citep{Nattila_2019}.
It employs the particle-in-cell (PIC) method with a standard 2nd-order field solver, charge-conserving current deposition scheme, and a relativistic Boris pusher.
The plasma is modeled with 8 particles per cell ($1728$ per $(c/\omega_{\mathrm{p}})^3$) and we smooth the electric current with 8 filter passes.
The time step is $0.45$ of the cell light-crossing time.
Further numerical details of the simulations are given in the Supplementary Materials.

\begin{figure}[t]
\centering
    \includegraphics[trim={0.2cm 0.4cm 0.4cm 1.2cm}, clip=true,width=0.45\textwidth]{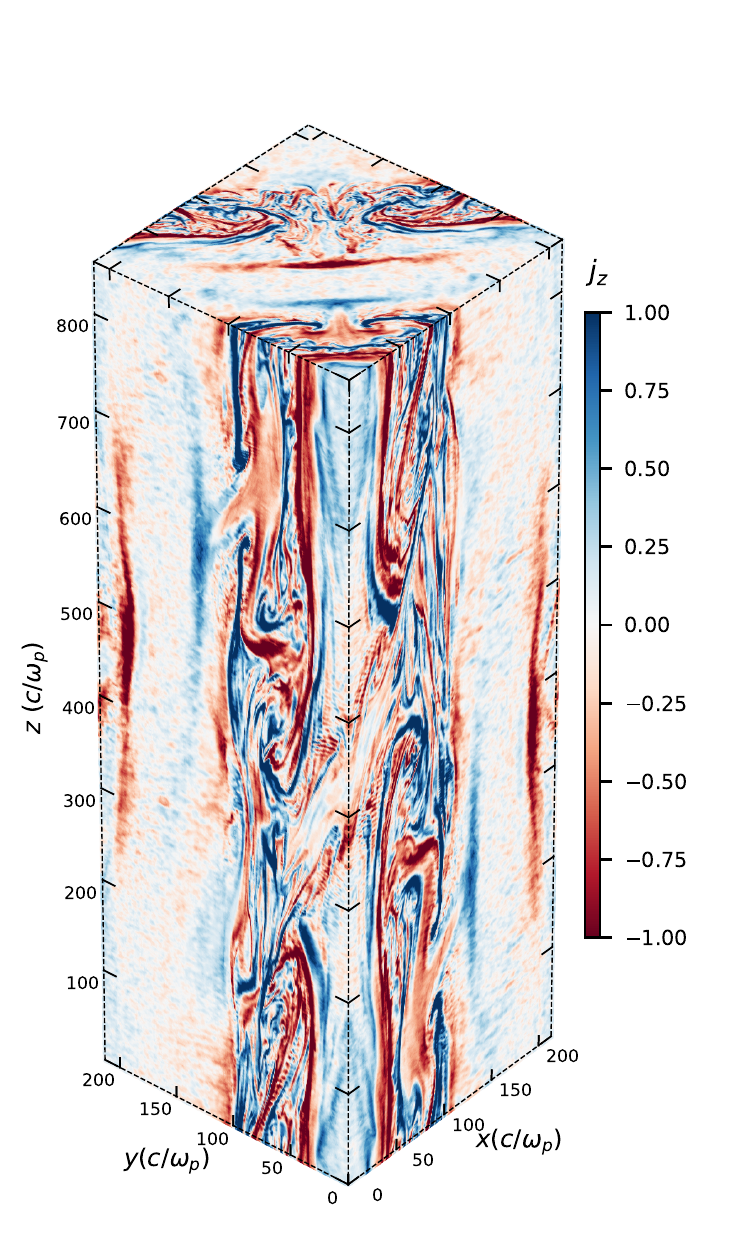}
    \caption{\label{fig:b05k2_3d}
3D visualization of the onset of the wave turbulence excited by the collision of two orthogonally polarized Alfv\'en waves with amplitudes of $\delta B/B_0 = 1/4$ and obliqueness $k_{\perp,0}/k_{\parallel,0} = 4$ at time $t/t_\parallel = 1.5$.
The magnetization parameter in this simulation is $\sigma_0=100$. 
Color shows the electric current density along the dominant magnetic field direction, $j_z$ (in units of $e n_0 c$).
Movie is available in Ref. \citep{Note1Movie}.
}
\end{figure}

\psection{Radiative ``ring-in-cell'' (RIC) simulations}
In addition to the PIC simulations, we use a novel RIC method, suitable for rapidly gyrating particles in strong magnetic fields, which tracks the particle's guiding center motion  without resolving the fast gyration \citep{Northrop_1963, Beklemishev_1999, Bacchini_2020, Note1GCA}.
This technique is particularly useful when simulating plasmas with $\sigma_0\gg 1$, where gyrofrequency $\omega_B=\sigma_0^{1/2}\omega_\mathrm{p}$ far exceeds the plasma frequency.
Strong magnetic fields often also imply fast synchrotron cooling. In particular, in neutron star magnetospheres 
the particles become confined to the ground Landau state and move along the magnetic field lines like beads on a wire.
This extreme limit can be simulated by damping any particle motion perpendicular to $\vec{B}$ (in $\vec{E}\times\vec{B}$ drift frame), which is easily implemented in a RIC simulation. 
Below we compare the results obtained in this extreme limit with the PIC simulation without synchrotron cooling. 
Remarkably, the wave and plasma evolution are very similar in the PIC and RIC simulations.

\psection{Nonlinear wave interactions}
We observed the following evolution.
The two initial, counterpropagating, orthogonally polarized waves deform the magnetic field lines effectively creating a sheared background for each other.
This causes the nonlinear interaction of the waves \citep{Howes_2013, Tenbarge_2021}. 
The Alfv\'en waves are sustained with a parallel electric current $\vec{j} \approx j\nvec{z}$.
As the waves pass through each other multiple times the currents become compressed into thinner sheets (Fig.~\ref{fig:b05k2_3d}).
This is the result of  nonlinear interaction producing daughter modes --- the thin sheets may be viewed as a superposition of the increasing number of high-frequency modes \citep{Howes_2016}.
This cascade process has previously been observed in gyrokinetic, MHD, and force-free electrodynamics simulations, and found responsible for the secular energy transfer to small scales \citep{Verniero_2018, Tenbarge_2021}.
The fact that our  kinetic simulations of relativistic turbulence show a similar evolution suggests that it is generic. 

Current sheet formation in MHD turbulence is presently a significant topic of research \citep{Mallet_2017, Mallet_2017a, Loureiro_2017, Boldyrev_2017, Ripperda_2021, Chernoglazov_2021}. 
Current sheets could become unstable to tearing, triggering magnetic reconnection. 
However, our simulation shows no signs of reconnection in the gradually thinning current sheets. 
As the evolution proceeds, we observe that the sheets develop more and more folds with increasing complexity and eventually become chaotic, filling the entire simulation domain.
An early phase of turbulence development is visualized in Fig.~\ref{fig:b05k2_3d}; a movie of the evolution is presented in Ref. \citep{Note1Movie}.
This nonlinear evolution involves primarily Alfv\'en waves with energy density $\sim (B_\perp^2 + E_\perp^2)/8\pi$. Compressive modes 
(in particular, perturbations of $B_z$ from the background $B_0$) develop at a lower level, contributing $\sim 10\%$ of the turbulence energy.

\begin{figure}
\centering
\includegraphics[width=0.48\textwidth, trim={0.0cm 0.0cm 0.0cm 0.0cm}, clip=true]{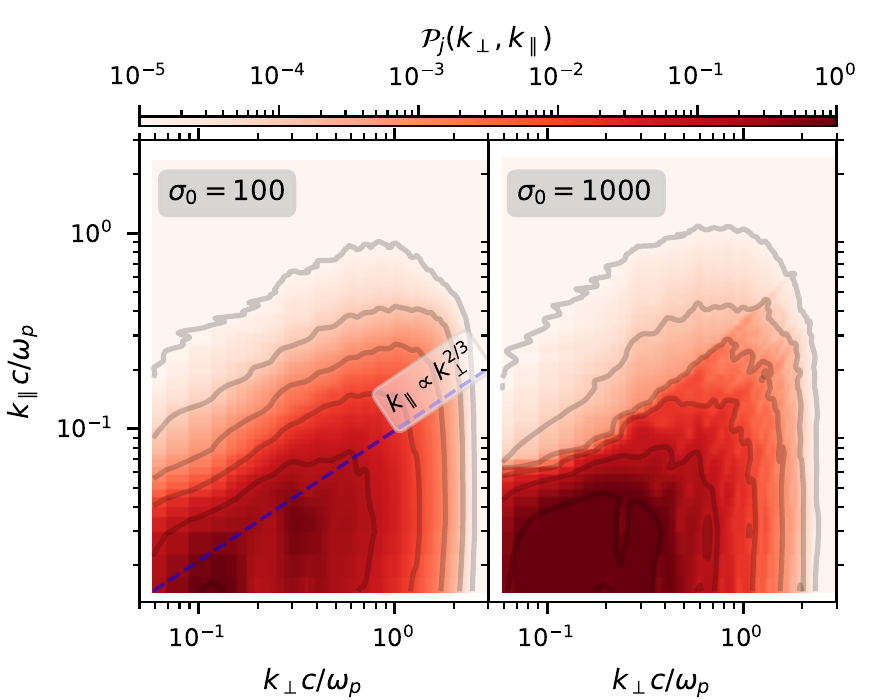}
    \caption{\label{fig:b05k1_spec}
        Power spectrum $\mathcal{P}_{j}(k_\perp,k_\parallel)$ of the electric current $j$ (in units of  $e n_0 c$), averaged over time window  $1<t/t_\parallel <1.5$, observed in two simulations with different $\sigma_0$ and the same fiducial parameters $\delta B/B_0=1/4$, $\chi_0=2$, and $\omega_\mathrm{p}/ck_{\perp,0}\approx 34$.
        The simulation with $\sigma_0=100$ (left) is charge surfeit, and the simulation with $\sigma_0=1000$ (right) is charge starved.
        The blue dashed line indicates the expected direction of the cascade development in the charge surfeit regime, $k_\parallel \propto k_\perp^{2/3}$.
}
\end{figure}

When viewed in Fourier space, the magnetic field and electric current fluctuations demonstrate the expected anisotropy: the cascade is $k_\perp$ dominated (Fig.~\ref{fig:b05k1_spec}).
The fluctuation spectrum in the $k_\perp,k_\parallel$ plane has practically no $k_\parallel$ dependence at $k_\parallel/k_{\parallel,0}<(k_\perp/k_{\perp,0})^{2/3}$ 
as expected for a critically-balanced cascade \citep{Goldreich_1995}.
The magnetic field power spectrum is similar but noisier than the electric current spectrum.

Besides the fiducial model with $\chi_0=2$ presented here, we ran other simulations with $1/16 \leq \delta B/B_0\leq 1/2$ and $1\leq k_{\perp,0}/k_{\parallel,0}\leq 8$. 
All of them excited turbulence cascades and showed a similar dissipation mechanism, independent of $\chi_0$. 
However, $\chi_0$ controls how quickly the cascade develops and becomes dissipative. 
 
\psection{Charge starvation}
The electric current density grows with $k_\perp$ in the cascade and may exceed $cen_0$. 
Then, the plasma becomes unable to support the high-frequency waves \citep{Blaes_1990, Thompson_2006a, Thompson_2008, Chen_2020}.
The parameter quantifying this ``charge-starvation'' effect for waves with amplitudes $B_\mathrm{w}(k_\perp)$ is 
\begin{equation}\label{eq:chargestarv}
    \kappa \equiv \frac{j}{e c n_0} 
           \approx  \frac{k_\perp B_\mathrm{w}}{4\pi e n_0}
           \approx \sigma_0^{1/2}\eta \left(\frac{ck_{\perp, 0}}{\omega_\mathrm{p}}\right)^{q}\left(\frac{ck_\perp}{\omega_\mathrm{p}}\right)^{1-q},
\end{equation}
where $\eta$ and $q$ parametrize the  magnetic field spectrum $B_\mathrm{w} = \eta B_0 (k_\perp/k_{\perp,0})^{-q}$.
One may expect the cascade to terminate because of charge starvation if $\kappa>1$ is reached at $k_\perp^{-1} > c/\omega_\mathrm{p}$.
This occurs if $\sigma_0 > \sigma_{\mathrm{cr}}=\eta^{-2}(c k_{\perp,0}/\omega_\mathrm{p})^{-2q}$. 
In the opposite case, the cascade is charge surfeit. 
Assuming a Kolmogorov-like turbulence spectrum ($q = 1/3$) our fiducial model parameters give $\sigma_{cr} \approx 170$ ($\sigma_{cr} \approx 530$ for weak turbulence with $q = 1/2$).
The effects of charge starvation may be studied by comparing the simulations with $\sigma_0=100$ (charge surfeit) and $1000$ (charge starved).

In the simulation with $\sigma_0=1000$ the magnetic field lines become stiffer, resisting sharp kinks that appeared in the charge-surfeit turbulence. 
The system tries to avoid starvation by increasing the local plasma density at the folds (by a factor $n/n_0=2-4$), however this does not prevent the suppression of turbulence at sufficiently high $k_\perp>k_{\perp,\mathrm{s}}$ at which $\kappa>1$.
Some of the cascading wave power is then redirected into expansion in the $\vec{k}$ space along $k_\parallel$ (see Fig.~\ref{fig:b05k1_spec}, right panel). 

The main consequence of charge starvation is the changed pattern of turbulence damping.
We observed that damping strongly varies on a timescale comparable to $\omega_\mathrm{s}^{-1}$ where $\omega_\mathrm{s} \approx c k_{\parallel,\mathrm{s}}$ is the frequency of the starving waves.
Damping peaks when the colliding waves are trying to produce  a daughter mode with $\omega>\omega_\mathrm{s}$ that is not supported by the plasma. 
This is accompanied by the development of a significant electric field component $E_\parallel$ (parallel to $\vec{B}$) comparable to the wave amplitude $B_\mathrm{w}$.

\begin{figure}
\centering
\includegraphics[width=0.48\textwidth]{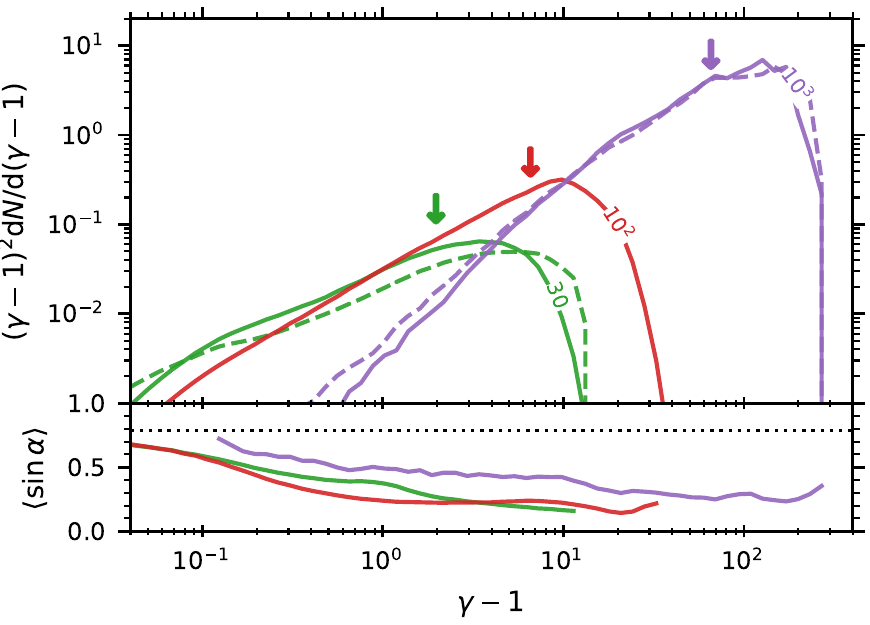}
    \caption{\label{fig:pspec}
        Particle kinetic energy spectra (top) and mean pitch angle (bottom) at a late time $t/t_\parallel = 3$ in the simulations with $\delta B/B_0=1/4$, $\chi_0 = 2$, and different 
        $\sigma_0=30$ (green), $100$ (red), and $1000$ (purple). 
        The fully kinetic PIC results are shown by the solid curves, and two RIC simulations for $\sigma_0=30$ and $1000$ are shown by the dashed curves. 
        Arrows indicate the mean  Lorentz factor at the end of turbulence dissipation, $\langle \gamma \rangle \approx (\delta B/B_0)^2 \sigma_0$.
        The horizontal dotted line in the bottom panel shows the mean pitch angle that would be found for an isotropic particle distribution.
}
\end{figure}

\psection{Particle energization}
In all the simulations, we find that damping of the Alfv\'en turbulence produces quasithermal heating of the plasma (see Fig.~\ref{fig:pspec}; top panel). 
There is no extended high-energy tail beyond the expected mean Lorentz factor $\langle \gamma \rangle \approx (\delta B/B_0)^2 \sigma_0$. 
This is in stark contrast with  simulations of turbulence with violent driving \citep{Zhdankin_2017a, Comisso_2018, Comisso_2019, Nattila_2021, Zhdankin_21, Hankla_21, Vega_21}.
As a result, all particles move mainly along $\vec{B}$.
The mean pitch angle
\citep{Note2}
of the hot particles is 2--4 times smaller than would be expected for an isotropic distribution (Fig.~\ref{fig:pspec},  bottom panel).
The small pitch angles $\sin\alpha\ll 1$ make the fully kinetic PIC simulations with no synchrotron cooling similar to the RIC simulations with complete synchrotron cooling.

The quasithermal heating along $\vec{B}$ demonstrates that the particles gain energy gradually, much slower than their gyration in the strong $\vec{B}$. 
This is true for both charge-surfeit and charge-starved cascades. 
With increasing scale separation $\omega_\mathrm{p}/ck_{\perp,0}$, which implies decreasing $B_\mathrm{w}/B_0$ on the damping scale, we expect the heating to become completely one-dimensional along $\vec{B}$. 

In the charge-surfeit regime, the expected heating mechanism is Landau damping. Alfv\'en waves with high $k_\perp$ (approaching the kinetic scale) have a significant $E_\parallel\sim B_\mathrm{w}$, because of the plasma inertia. 
Their phase speed along $\vec{B}$ drops below $c$ \citep{Arons_1986}, so particles can resonantly exchange energy with the high-$k_\perp$ waves.
A similar damping mechanism has been invoked for the nonrelativistic turbulence in the solar wind  \citep{TenBarge_2013a, Li_2016a, Howes_2018, Chen_2019}.
By contrast, in the charge-starved regime $E_\parallel\sim B_\mathrm{w}$ develops in waves with $k_\perp\sim k_{\perp,\mathrm{s}}$, which have $\kappa\sim 1$. 

In both regimes, particle acceleration is slower than gyration. 
Furthermore, bending of the magnetic field lines at the damping scale is small, $B_\mathrm{w}/B_0\ll 1$. 
As a result, particle motion perpendicular to $\vec{B}$ cannot be efficiently excited, and almost all the turbulence energy converts into particle motion along $\vec{B}$.
We have measured the dissipation rate $W_\parallel=\langle\vec{E}_\parallel \cdot \vec{j}\rangle$ (averaged over space and time) and compared it with $W_\perp=\langle \vec{E}_\perp \cdot \vec{j}\rangle$ where $\vec{E}_\perp$ is the electric field component perpendicular to $\vec{B}$. 
We found $W_\parallel/W_\perp \approx 6.4$ for $\sigma_0 = 30$ and $W_\parallel/W_\perp \approx 16$ for $\sigma_0 = 1000$, confirming the dominant role of $E_\parallel$.

\psection{Discussion}
Our simulations are among the largest 3D kinetic simulations of relativistic turbulence, reaching the scale separation $\omega_\mathrm{p}/c k_0\approx 34$ in our fiducial runs with the high resolution of the dissipation scale, and up to $\omega_\mathrm{p}/c k_0=170$ in shorter runs employing domains of $2560^2 \times 10\,240$ and a reduced resolution of $c/\omega_\mathrm{p}$.
The simulations may still be unable to capture possible effects special to cascades with extremely large inertial ranges. 
In particular, it was argued that Alfv\'enic cascades with very large scale separations may develop tearing-unstable current sheets, releasing magnetic energy via magnetic reconnection \citep{Mallet_2017, Mallet_2017a, Loureiro_2017, Boldyrev_2017, Ripperda_2021}.
If reconnection is activated on small scales and dissipates field jumps $\Delta B\ll B_0$, it would kick particles to modest $\gamma \sim (\Delta B/B_0)^2 \sigma_0$. Such dissipation in the dominant guide field would be unable to ``unmagnetize'' the particles --- their motion would remain confined along $\vec{B}$ (resembling particle energization we observed in the charge-starved cascade). 

Deviations from the picture described in this Letter could occur if the turbulence is excited with a large $\delta B/B_0\gtrsim 1$, depending on how it is driven.
We have also simulated collisions of Alfv\'en-wave packets with $\delta B/B_0 >1$ and experimented with both plane and torsional waves (see Refs. \citep{Li_2019, Ripperda_2021} for similar setups).
We observed that such collisions immediately emit a large fraction of the wave energy in a strong compressive ``fast'' mode --- the second eigen mode of the magnetically dominated plasma, which can freely propagate across $\vec{B}$ and escape the system. 
The remaining Alfv\'en waves developed the turbulence cascade resembling our simulations with lower $\delta B/B_0$.  

Violent deformation of the magnetic field $\delta B/B_0 \gtrsim 1$ with an arbitrary perturbation pattern may qualitatively differ from proper Alfv\'enic turbulence. 
Then, immediate formation of large tearing-unstable current sheets, was observed, followed by strong nonthermal particle acceleration \citep{Comisso_2018, Comisso_2019, Nattila_2021}. 
Unlike Alfv\'en wave turbulence cascades, this violent dissipation occurs in a similar way in 2D and 3D simulations.
Such reconnection-mediated energy release may result from global instabilities triggered by overtwisting of magnetic field lines, and it should be distinguished from damping of Alfv\'enic turbulence cascades in a dominant background field $\vec{B}_0$. 

This distinction has important observational implications.
For Alfv\'enic turbulence, our simulations show the absence of nonthermal particle acceleration, and heating occurs along $\vec{B}$.
Then, synchrotron emission will be suppressed, and the dominant radiative process is inverse Compton (IC) scattering of ambient photons, which produces hard radiation spectra (see also Refs. \citep{Sobacchi_2021a, Sobacchi_2021b, Comisso_2021a, Comisso_2021b}).
Synchrotron emission from plasmas heated by Alfv\'enic turbulence can be activated if the IC photons do not escape the system and turn into secondary $e^\pm$ pairs, after a sufficient free path that allows them to develop pitch angles relative to local $\bB$ \citep{Beloborodov_2017, Sobacchi_2021b}.
Copious pair creation will occur in sufficiently compact systems, and can even make the plasma optically thick. 
In this high-compactness regime, the radiative output of Alfv\'enic turbulence can become similar to flares emitted by large-scale magnetic reconnection \citep{Beloborodov_2021}.
For less compact, optically thin systems, one can expect a drastic difference between damping of Alfv\'enic turbulence  and large-scale magnetic reconnection: 
plasma heated by Alfv\'enic turbulence will be synchrotron silent. 
This difference may help identify the energy release mechanism in observed sources. 

\begin{acknowledgments}
\psection{Acknowledgments}
We thank
Stanislav Boldyrev,
Luca Comisso, 
Daniel Gro\v{s}elj, 
Yuri Lyubarsky,
Jens Mahlmann,
Alfred Mallet,
Donald Melrose,
Sasha Philippov, 
Bart Ripperda, 
Alex Schekochihin,
Lorenzo Sironi, 
Emanuele Sobacchi,
Chris Thompson,
and the anonymous referees for helpful discussions and comments. 
J.N. is supported by a joint Columbia University/Flatiron Research Fellowship.
A.M.B. is supported by NSF Grants No. AST~1816484 and No. AST~2009453, Simons Foundation Grant No. \#446228, and the Humboldt Foundation.
The computational resources and services used in this work were provided by facilities supported by the Scientific Computing Core at the Flatiron Institute, a division of the Simons Foundation.

\end{acknowledgments} 

\bibliographystyle{apsrev}

\begin{thebibliography}{43}
\bibitem[{Note1GCA()}]{Note1GCA}
\bibinfo{note}{Our implementation includes only acceleration by
  $\protect \mathbf {E}_\parallel $ and the $\protect \mathbf {E}\times
  \protect \mathbf {B}$ drift term when updating the particle momentum. Higher
  order drift terms are found unimportant for simulations with $\delta B/B_0 <
  1$.}
\end{thebibliography}

\newpage
\begin{appendix}
\section{Supplementary Material}

\subsection{A: Technical Parameters and Numerical Tests}

Our simulations are performed with the open-source, massively-parallel, particle-in-cell code \textsc{runko} \citep{Nattila_2019} using a public release version \textsc{v3.1} (Ripe Mango; GitHub repository commit \textsc{137b81a}).
The simulations employ the second-order field solver (\textsc{FDTD2}), relativistic Boris pusher (\textsc{BorisPusher}) or the ring-in-cell pusher (reduced guiding center pusher;  \textsc{rGCAPusher}), charge-conserving current deposition scheme (\textsc{ZigZag}), linear field interpolator (\textsc{LinearInterpolator}),
and Binomial 3-point digital filter (\textsc{Binomial2}).

One source of numerical noise in PIC simulations is the finite number of plasma particles.
We performed test simulations where pair plasma is initialized with 8 and 32 particles per cell, and found that 8 was sufficient, as this 
gave practically the same results 
(same particle spectra and turbulence dissipation rates)
as with 32 particles per cell. 
Therefore, to reduce computational costs, our largest and longest fiducial models use 8 particles per cell. 

Particles in PIC simulations are clouds of a finite size, which ``deposit'' electric current to the grid.   
The level of the high-frequency numerical noise in the deposited current depends on  the cloud shape function.  
We use the computationally cheap first-order shape function, and suppress the high-frequency noise using 8 passes of digital filter, smoothing the electric current before it is deposited on the field grid.
We have ran identical simulations without any current filtering and verified that  they also reproduce the turbulence dynamics and particle energization rates, although with a stronger numerical noise.

Another source of numerical error can be the finite grid resolution.
The grid must resolve the smallest scale of kinetic plasma phenomena, which is the skin-depth $c/\omega_\mathrm{p}$ for relativistic plasma.
Our fiducial simulations resolve the plasma skin depth with 6 cells.
We have verified convergence by running test simulations with up to 12 cells per skin depth. 

We have also studied the role of the plasma initial state by initializing simulations with different momentum distributions of plasma particles. 
In particular, we verified that choosing an extremely low temperature $\theta_0\equiv k_\mathrm{B} T_0/m_\mathrm{e} c^2=10^{-5}$ (which implies an initially unresolved Debye length) gives the same results as simulations with $\theta_0=0.3$ (for which the Debye length is well resolved from the beginning). 
We observed that the choice of $\theta_0$ is unimportant, as long as $\theta_0$ is below temperatures produced by the turbulent heating, $\theta_0 \ll (\delta B/B_0)^2 \sigma_0$.

\end{appendix}

\end{document}